\documentstyle[aps,psfig,amsmath]{revtex}
\newcommand{\be}{\begin{equation}}
\newcommand{\ee}{\end{equation}}
\newcommand{\bea}{\begin{eqnarray}}
\newcommand{\eea}{\end{eqnarray}}
\newcommand{\bef}{\begin{figure}}
\newcommand{\eef}{\end{figure}}
\newcommand{\bm}{\bibitem}

\newcommand{\bt}{\beta}
\newcommand{\gm}{\gamma}
\newcommand{\th}{\theta}

\newcommand{\sg}{\sigma}
\newcommand{\de}{\delta}
\newcommand{\ep}{\epsilon}
\newcommand{\rw}{\rightarrow}

\newcommand{\cl}{{\cal{L}}}

\newcommand{\bp}{{\boldsymbol{p}}}
\newcommand{\bsg}{{\boldsymbol{\sigma}}}
\newcommand{\bl}{{\boldsymbol{l}}}
\newcommand{\bbm}{{\boldsymbol{m}}}
\newcommand{\bn}{{\boldsymbol{n}}}

\newcommand{\ov}{\overline}
\newcommand{\oM}{\overline M}
\newcommand{\of}{\overline f}
\newcommand{\on}{\overline n}

\newcommand{\bu}{\ov {u}}
\newcommand{\la}{\langle}
\newcommand{\ra}{\rangle}
\newcommand{\ps}{p\!\!\!/}

\begin{document}

\setcounter{page}{1}
                                                                                
\title{On the nucleon self-energy in nuclear matter}

\author{S. Mallik}
\address{Saha Institute of Nuclear Physics, 1/AF, Bidhannagar, 
Kolkata-700064, India}

\author{Andreas Nyf\/feler}
\address{Institute for Theoretical Physics, ETH, CH-8093, Z$\ddot{u}$rich, Switzerland}

\author{M. C. M. Rentmeester}
\address{Institute for Theoretical Physics, University of Nijmegen,
Nijmegen, The Netherlands}

\author{Sourav Sarkar}
\address{Variable Energy Cyclotron Centre, 1/AF, Bidhannagar, 
Kolkata-700064, India}

\maketitle
                                                                                
\begin{abstract}

We consider the nucleon self-energy in nuclear matter in the absence of
Pauli blocking. It is evaluated  using the partial-wave analysis of $N\!N$ 
scattering data. Our results are compared with that of a realistic
calculation to estimate the effect of this blocking. It is also possible to 
use our results as a check on the realistic calculations.    

\end{abstract}

\section{Introduction}

A fundamental problem of nuclear physics is to explain the properties of 
nuclear matter (and finite nuclei) in terms of an effective field theory at 
low energy based only on the (chiral) symmetry of QCD. While such a theory 
has been eminently successful for systems like $\pi\pi$ and $\pi N$
\cite{Ecker}, a satisfactory theory for the $N\!N$ system has been difficult 
to formulate due to the presence of two-nucleon bound or virtual states close 
to the threshold of $N\!N$ scattering \cite{Weinberg1}. In particular, the
leading chiral four-nucleon interaction predicts an absurdly large value for 
the self-energy of the nucleon at normal nuclear density \cite{Montano}. 

There is, however, a semi-phenomenological approach that yields fairly
accurate values for different observables in nuclear matter. Here the 
$N\!N$ potential is constructed by exchanging low mass bosons in the 
$t$-channel \cite{Potentials}. The coupling and other parameters in the 
potential are determined by experimental data on the deuteron and the low 
energy $N\!N$ scattering. The dynamics is formulated on the basis of a
relativistic version of the Brueckner-Hartree-Fock method 
\cite{Anastasio,Brockmann}, where the reaction matrix satisfies a 
three-dimensionally reduced Bethe-Salpeter equation in nuclear medium. 
The Dirac equation for the nucleon incorporates the scalar part of the 
self-energy due to its interaction with nucleons in the medium. The
self-energy itself is given by the diagonal element of the reaction matrix. 
The system of equations is then solved self-consistently.

In this work we study the nucleon self-energy in a certain theoretical
limit. We observe that if we suspend the Pauli blocking operator 
(projecting onto the unoccupied states) in the equation for the reaction 
matrix, it coincides with the one for the scattering matrix in vacuum. 
Further, if we do not include the relevant part of the single particle 
self-energy in the mass term in the Dirac equation, the requirement of 
self-consistency does not arise any more. In this limit the self-energy 
is given by an integral over the spin-averaged, forward $N\!N$ scattering 
amplitude in vacuum, which can be evaluated entirely with the experimental
data. 

There is a well-known expansion in statistical mechanics, called the virial
expansion, whose first term for the in-medium self-energy would give 
exactly the theoretical limit considered above \cite{Leutwyler,Jeon,Mallik}. 
Here we employ this method to derive the formula for the limiting 
self-energy of the nucleon. We then evaluate it at different nuclear 
densities, using the phase-shift analysis of $N\!N$ scattering data, 
independently of any $N\!N$ potential. It is then compared with the 
realistic calculation \cite{Brockmann} to assess the importance of the 
effect of Pauli blocking in nuclear matter.

Our calculation would also serve as an important check on the realistic
calculation. One has just to repeat the calculation of the self-energy  
in the original framework itself, using the phenomenological potential 
and the physical nucleon mass, but in the theoretical limit of omitting 
the Pauli blocking operator in the reaction matrix. This result as a 
function of the nuclear density must agree with that of the present 
calculation. This, in turn, would confirm our assertion that it is indeed
the Pauli blocking effect which distinguishes the realistic calculation 
from the one presented here.
 
\section{Derivation of self-energy formula}
\setcounter{equation}{0}
\renewcommand{\theequation}{2.\arabic{equation}}

Here we obtain the leading term in the virial expansion for the nucleon
self-energy in nuclear medium \cite{Leutwyler,Jeon,Mallik}. We begin by
stating clearly the normalization of different
quantities. Omitting the nucleon isospin index \cite{Comment1}, we take the 
creation and the annihilation operators for the nucleon with momentum $\bp$ 
and spin projection $\sg\,(=\pm\frac{1}{2})$ to satisfy the anticommutation 
relation,
\be
\{b(\bp,\sg),b^\dag(\bp',\sg')\}=(2\pi)^3\,2E_p\,\de(\bp-\bp')\de_{\sg\sg'}\,
, ~~~  E_p=\sqrt{\bp^2+m^2}\,.
\ee
The single nucleon state is defined as
$|\bp,\sg\ra=b^\dag(\bp,\sg)|0\ra~$. Similarly the two nucleon state is 
$|\bp_1,\sg_1;\bp_2,\sg_2\ra=b^\dag(\bp_1,\sg_1)b^\dag(\bp_2,\sg_2)|0\ra$.
Clearly their normalization is fixed by the anticommutation rule (2.1). 
The (positive energy) Dirac spinors are normalized such that the spin sum
over these spinors, to be used below, is given by
\[\sum_\sg u(\bp,\sg)\bu(\bp,\sg)=\ps+m~.\]

The derivation starts by considering the nucleon self-energy in 
{\it vacuum}, which may be expressed as an $S$-matrix element,
\be
-i(2\pi)^4\de^4(p'_1-p_1)\bu(\bp'_1,\sg'_1)\Sigma^{(0)}(p)u(\bp_1,\sg_1)=
\la0|b(\bp'_1,\sg'_1)\,(S-1)\,b^\dag(\bp_1,\sg_1)|0\ra~,
\ee
where $S$ is the familiar scattering matrix operator, 
$S=Te^{i\int\cl_{int}(x)\,d^4x}~,$ for an interaction
Lagrangian $\cl_{int}$.
The subscript $1$ on the variables of the particle anticipates another one 
with which it will interact in the medium. In fact, we shall express below 
the nucleon self-energy in {\it nuclear medium} in terms of the $N\!N$ 
scattering amplitude in {\it vacuum}, defined as usual by
\be
\la\bp'_1,\sg'_1;\bp'_2,\sg'_2|S-1|\bp_1,\sg_1;\bp_2,\sg_2\ra=
i(2\pi)^4\de^4(p'_1+p'_2 -p_1-p_2)\,M(p_1,\sg_1;\,p_2,\sg_2\rw 
p'_1,\sg'_1,; p'_2,\sg'_2)\, ,
\ee
where $M$ stands for the scattering matrix sandwiched between spinors
corresponding to the final and the initial states of the two nucleons. 

Below we shall meet the spin averaged amplitude in the forward direction,
\be
{\ov M}(p_1,p_2\rw p_1,p_2))=\frac{1}{4}\sum_{\sg_1,\sg_2}\,
M(p_1,\sg_1;p_2,\sg_2\rw p_1,\sg_1;p_2,\sg_2),
\ee
With our normalization of states, the amplitude $\oM$ is Lorentz invariant.

To obtain the self-energy in nuclear medium, we have to replace the vacuum
expectation value in Eq.~(2.2) by an appropriate one. Although we specialize
later to zero temperature, we take here the most general average over an 
ensemble of systems maintained at temperature $T (=1/\bt)$ with nucleon chemical
potential $\mu$. Thus the in-medium self-energy $\Sigma$ is given by 
\be
-i(2\pi)^4\de^4(p'_1-p_1)\bu(\bp'_1,\sg'_1)\Sigma(p)u(\bp_1,\sg_1)=
\la b(\bp'_1,\sg'_1)(S-1)b(\bp_1,\sg_1)\ra~,
\ee
where for any operator $O$,
\[\la O\ra=Tr[e^{-\bt(H-\mu {\cal N})}O]/Tr e^{-\bt(H-\mu{\cal N})}~.\]
Here $H$ is the Hamiltonian and ${\cal N}$ the nucleon number density 
operator. Clearly this form of the Boltzmann weight breaks explicit Lorentz 
invariance and singles out the rest frame of the medium \cite{Comment2}. 

We now make use of the virial expansion to first order in density. For an
operator $O$, the ensemble average in nuclear medium can be expanded as 
\[\la O\ra=\la 0|O|0\ra+\sum_{\sg_2}\int\frac{d^3p_2}{(2\pi)^3
2E_{p_2}}n(E_{p_2})
\la\bp_2,\sg_2|O|\bp_2,\sg_2\ra~,\]
where $n(E_p)$ is the nucleon distribution function, 
$n(E_p)=1/[e^{\beta (E_p-\mu)} +1]$.
Applying it to the left hand side of Eq.~(2.5), we get for the difference
$\Sigma^{(n)}(p)=\Sigma(p)-\Sigma^{(0)}(p),$
\be
-i(2\pi)^4\de^4(p'_1-p_1)\bu(\bp'_1,\sg'_1)\Sigma^{(n)}(p_1)u(\bp_1,\sg_1)=
\sum_{\sg_2}\int\frac{d^3p_2}{(2\pi)^3 2E_{p_2}}n(E_{p_2})
\la\bp_2,\sg_2|b(\bp'_1,\sg'_1)(S-1)b^\dag(\bp_1,\sg_1)|\bp_2,\sg_2\ra~.
\ee
The matrix element in Eq.~(2.6) will be immediately recognised to be
the $N\!N$ scattering amplitude defined above by Eq.~(2.3). Cancelling the 
$\de$-function on both sides, we set $\sg'_1=\sg_1$ and sum over $\sg_1$
also to get
\be
-tr \{\Sigma^{(n)}(p_1)(\ps_1+m)\}=4\int\frac{d^3p_2}{(2\pi)^3
2E_{p_2}}n(E_{p_2}) \oM(p_1,p_2\rw p_1,p_2),
\ee
where the tr(ace) is over matrices in Dirac space. Note the similarity of
this equation with the corresponding one in Brueckner theory
\cite{Feshbach}. There is, however, one important difference: Our first
order formula has the scattering amplitude {\it in vacuum}, while it is the
amplitude {\it in medium} that enters the equation in Brueckner theory.
We shall discuss this point again in Sec.IV.

So far we did not state explicitly the isospin structure of the amplitude
$\oM$, which is now easy to figure out. We consider 
symmetric nuclear matter and work in the limit of iso-spin symmetry. Let the 
traversing nucleon be in any one of its isospin states, say a proton.
It may scatter with a proton or a neutron in the medium. The amplitude is
therefore given by the sum,
\be
\oM=\oM_{pp\rw pp}+\oM_{pn\rw pn}~. 
\ee

We now restrict to the case, where the three-momentum  $\bp_1$ is set equal
to zero. Then the rest frame of the medium coincides with the lab frame of 
the scattering process. The self-energy in this frame has the simple Dirac 
matrix structure,
\be 
\Sigma^{(n)}=U\cdot 1+V\gm^0,
\ee 
where the coefficients $U$ and $V$ depend only on the nucleon density.
Then the left hand side of Eq. (2.7) simplifies to $-4m(U+V)$.
On the other hand, the nucleon propagator with self-energy correction,
$i/\{\ps_1-m-\Sigma^{(n)}(p_1)\}$, reduces, for $\bp_1=0$, to
\be
\frac{i}{p_0-(m+U+V)}\frac{1}{2}(1+\gm^0)~,
\ee
in the vicinity of the pole. The shifted pole position is thus given by
\be
m^*-\frac{i}{2}\gm = m+U+V = m 
-\frac{1}{m}\int\frac{d^3p_2}{(2\pi)^3 2E_{p_2}}n(p_2) \oM(p_2),
\ee
where $m^*$ is the effective mass of the nucleon and $\gm$ gives the damping
rate of nucleonic excitations.

\section{Evaluation}
\setcounter{equation}{0}
\renewcommand{\theequation}{3.\arabic{equation}}

In our evaluation we restrict ourselves to nuclear matter at
zero temperature. In this limit we assume the nuclear medium to be a
non-interacting Fermi gas with all states filled upto the Fermi momentum
$p_F$, so that $n(\bp)\rw \th (p_F-|\bp|)$. For the symmetric medium the number 
density is then given by
\[ \on= 4\int \frac{d^3p}{(2\pi)^3}\th (p_F -|\bp|)
=\frac{2p_F^3}{3\pi^2}~,\]
where $p_F$ is related to the chemical potential by  $p_F=\sqrt{\mu^2-m^2}$.

The scattering amplitudes are generally analysed in the center-of-mass 
(c.m.) frame, where they are normalized in a slightly different way to get 
a simple expression for the differential cross section. One defines a 
scattering amplitude $f$ related to $M$ by
\be
\frac{d\sg}{d\Omega}=\frac{| M|^2}{(8\pi E)^2}\equiv | f|^2
\ee
where $E$ is the total energy in the c.m. frame. The results of partial wave
analysis are generally given as functions of the lab kinetic energy 
$T (=\sqrt{p_2^2+m^2} -m)$, in terms of which we have $E=\sqrt{2m(2m+T)}$. 
We also note here the expression for the c.m. momentum in terms of $T$, 
$k=\sqrt{mT/2}$.

We can now write Eqs. (2.11) as 
\be
m^*-\frac{i}{2}\gm =m-\frac{2}{\pi}\sqrt{\frac{2}{m}} \int_0^{T_F} dT
\sqrt{T}(2m+T){\ov f}(T)~,
\ee
where $T_F$, the upper limit of the integral, is related to $p_F$ by 
$p_F= \sqrt{T_F(2m+T_F)}$ and 
\[\of=\of_{pp \rw pp} +\of_{pn \rw pn} ,\]
the bar indicating spin averaging as in Eq. (2.4). In terms of amplitudes
with definite iso-spin,
\[\of =3/2 \of ^{(I=1)} +1/2 \of ^{(I=0)} .\]

The full scattering amplitudes are expanded in a series of partial wave 
amplitudes, which may then be determined by fitting with experimental 
scattering data. For the spin averaged, forward scattering amplitudes,
this expansion takes a particularly simple form \cite{Bohr,Weinberg2},
\be
{\of}^I (E)= 2\cdot\frac{1}{4}\sum_{jsl}(2j+1) f^{Ijs}_{l,l} (E)
\ee
Here the factor of 2 takes into account the identity of the two nucleons 
in the scattering process. The total angular momentum $j$ is obtained by 
coupling the total spin and orbital angular momenta $s$ and $l$ respectively. 
The Pauli principle restricts the possible amplitudes by requiring the
quantum numbers to satisfy
\[(-1)^l (-1)^{1-s} (-1)^{1-I} =-1 \]
In general, the amplitude $f^{Ijs} (E)$ is a matrix in the $l$ space, whose
diagonal elements enter the sum in Eq.~(3.3). Below we shall remove the
superscripts $I$ and $s$ on the partial wave amplitudes.

The form of the partial wave amplitudes is determined by the unitarity of
the $S$-matrix. Thus for the uncoupled waves, $s=0, l=j$ and $s=1, l=j$, we
have the single element,
\[f^j_{l,l} (E)\equiv f^j_l(E)=\left( e^{2i\de^j_l (E)}-1\right)/2ik, ~~~l=j\,, \]
where $\de^j_l$ is the phase shift, a real function of $E$. But the 
waves $s=1, l=j\pm 1$ are coupled, leading to a $2\times 2$ matrix amplitude
with the diagonal elements of the form \cite{MacGregor},
\[f^j_{l,l} (E)=\left( e^{2i\de_l^j (E)} cos \ep_j(E) -1\right)/2ik, ~~~l=j\pm 1. \]
where we have the mixing parameters $\ep_j(E) $ in addition to the phase
shifts.

We may now evaluate the integral (3.2), using the phase shift analysis of 
the Nijmegen group \cite{Nijmegen}. Alternatively we may take advantage of 
the reconstruction of the full (Saclay) amplitudes from this analysis, also 
carried out by the same group. The general amplitude may be written in the 
c.m. frame as a $4\times 4$ matrix in the Pauli basis as \cite{Winternitz}
\be
{\cal M}(\bp,\bp')=\frac{1}{2}\{a_s+b_s
+(a_s-b_s)\bsg_1\!\cdot\!\bn \,\bsg_2\!\cdot\!\bn
+(c_s+d_s)\bsg_1\!\cdot\!\bbm\,\bsg_2\!\cdot\!\bbm
+(c_s-d_s)\bsg_1\!\cdot\! \bl\, \bsg_2\!\cdot\!\bl\}
\ee
where the Saclay amplitudes, $a_s,\,b_s\,,c_s$ and $d_s$, are
complex functions of the energy and scattering angle. 
(We omit a fifth amplitude, which is zero in the forward direction.) 
Here $\bl\,, \bbm$ and $\bn$ are three mutually orthogonal unit vectors. 
The Pauli matrices $\bsg_1\,, \bsg_2$ act on the Pauli spinor $\chi$'s of 
the first and second nucleon. The spin-averaged, forward amplitude ${\ov f}$ 
is obtained from $\cal M$ as
\be
{\ov f} (E)=\frac{1}{4}\sum_{\sg_1,\sg_2}
\chi^\dag_{\sg_2}\chi^\dag_{\sg_1}{\cal M}(\bp,\bp)\chi_{\sg_1}\chi_{\sg_2}
=\frac{1}{2} (a_s(E) +b_s(E))
\ee
With the values of the Saclay amplitudes \cite{Nijmegen}, we evaluate the 
real and the imaginary parts of the integral (3.2) at different Fermi
momenta. The results are shown by the solid curves in Figs. 1 and 2.

\bef
\centerline{\psfig{figure=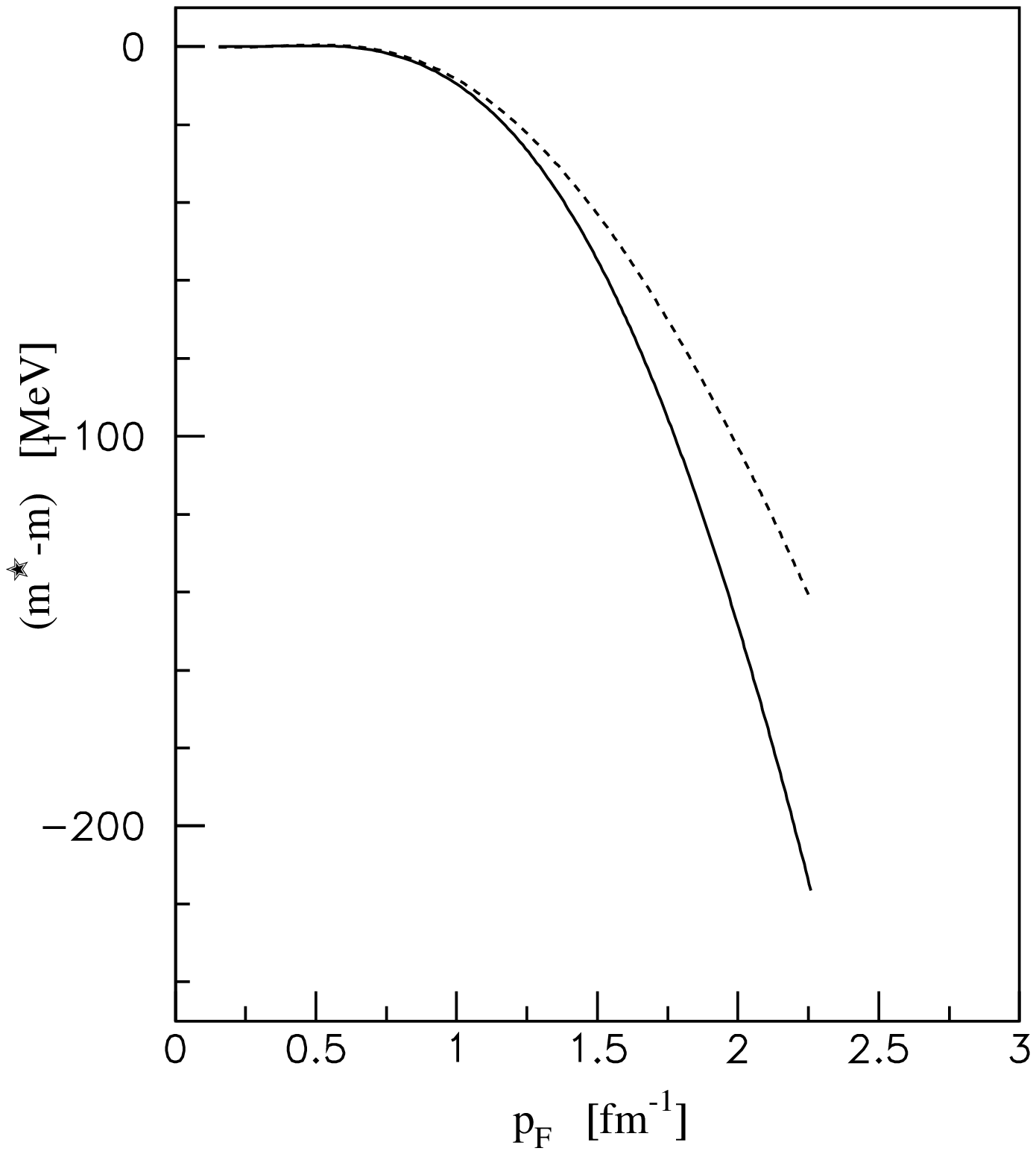,height=7cm,width=9.5cm}}
\caption{Shift in nucleon mass in nuclear matter as a function of Fermi
momentum. The solid curve results from the partial-wave analysis of the
Nijmegen group, while the dashed one is from the s-waves in
the effective range approximation.}
\eef

\bef
\centerline{\psfig{figure=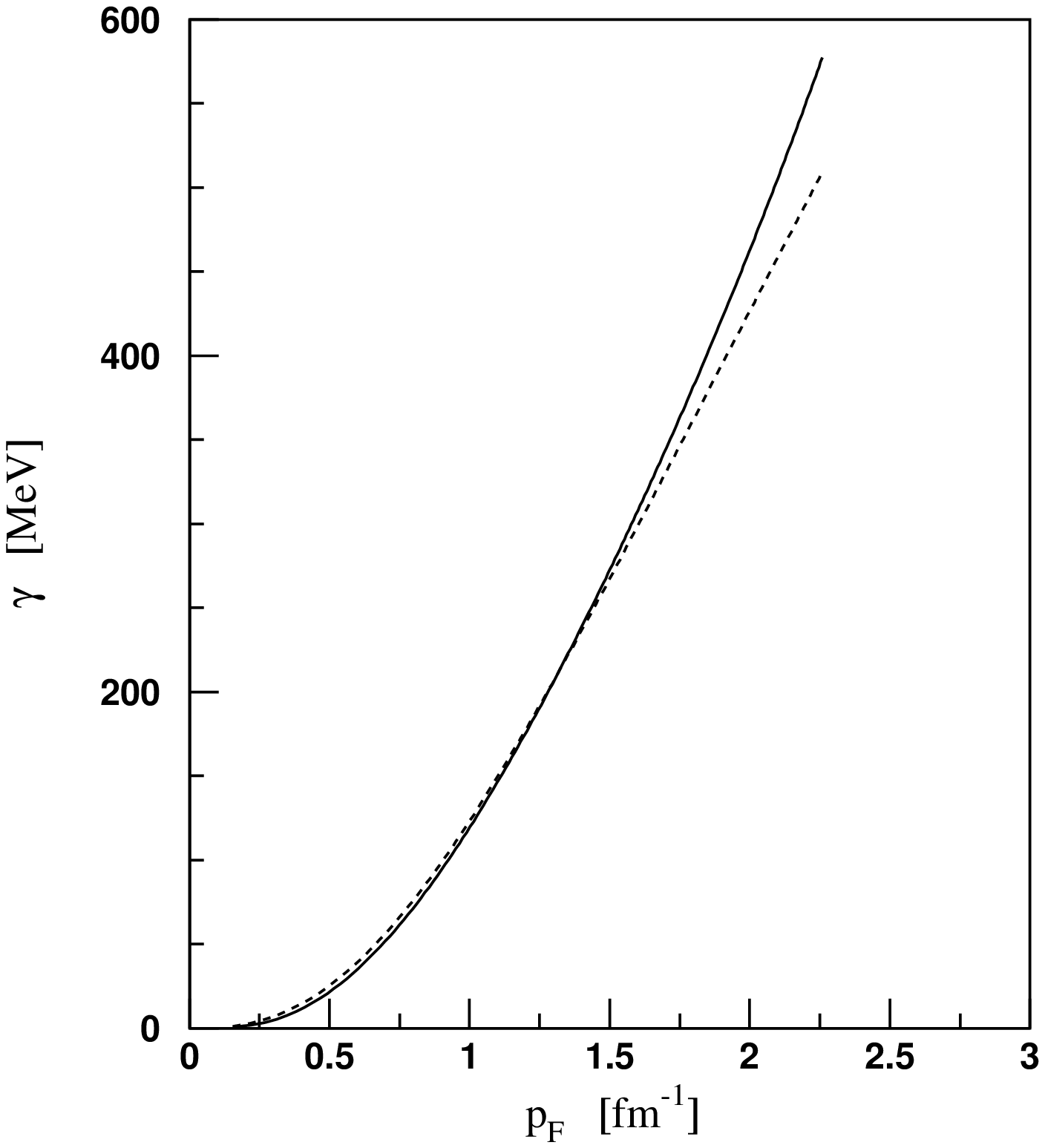,height=7cm,width=9.5cm}}
\caption{Damping rate of nucleonic excitation in nuclear matter 
as a function of Fermi momentum. The origin of solid and dashed curves 
are the same as in Fig.1.}
\eef

For an independent, but approximate, estimate, we also evaluate the integrals 
by including only the s-waves in the effective range approximation. Here an
s-wave amplitude is written as $f_0=1/(k\,cot\de -ik)$ with 
$k\,cot\de=-a^{-1}+rk^2/2$, where $a$ and $r$ are the scattering
length and the effective range. The values of these constants are long
known \cite{Bohr}: For the spin singlet state, $\, a =-23.7,\; r=2.7$ and for
the spin triplet state, 
$\, a=5.39,\; r=1.70 $, all in units of fm. This evaluation is shown by the 
dotted curves in Figs.1 and 2. It is seen that the higher partial waves 
contribute little up to about $p_F=1 fm^{-1}$. 

\section{Discussion}

Here we have considered the nucleon self-energy in nuclear matter, 
in the limit of ignoring the effect of Pauli blocking on it. This
self-energy can be expressed in terms of the forward spin-averaged $N\!N$ 
scattering amplitude in vacuum. We calculate its real and imaginary parts,
using the phase-shift analysis of experimental data on $N\!N$ scattering.

Our results may be compared with that of the self-consistent Hartree-Fock
calclation  \cite{Brockmann} to get an idea of the importance of Pauli 
blocking in the Fermi medium. (Since we do not include the relevant part 
of the self-energy in the nucleon mass, it would be appropriate to compare 
our results with their so-called 'non-relativistic' version of results.) 
In both the calculations the mass-shift is a strongly dependent function 
of the nuclear density. At normal nuclear density ($p_F=1.35 \, fm^{-1}$) 
they find the (real part of the) mass shift to be $-87$ MeV; in our 
calculation this value is attained at a higher density corresponding to 
$p_F=1.70 \, fm^{-1}$.

Our calculation, which is based only on experimental data, readily provides
a check on the the original relativistic Brueckner calculation
\cite{Brockmann}. We just need redo  this calculation without the Pauli 
operator in the equation for the reaction matrix. The resulting calculation 
with the phenomenological potential should yield the same functional
dependence of the self-energy on nuclear density as we find here.

\section*{acknowledgements}

One of us (S.M.) thanks Prof. A. Harindranath for discussions and Prof. M.G. 
Mustafa for help in preparing the manuscript. He also acknowledges the support 
of CSIR, Government of India.

\end{document}